\documentclass[aps,twocolumn,groupedaddress,showpacs]{revtex4}%
\usepackage{amsmath}
\usepackage{graphicx}%
\usepackage{amsfonts}%
\usepackage{amssymb}
\usepackage{epsf}

\begin{document}
\bibliographystyle{apsrev}

\title{Boundary Poisson structure and quantization}
% Optional argument for running titles on pages
%\title[]{}


\author{Liu Zhao}% \hspace{1cm}
%\email{lzhao@nwu.edu.cn}
\affiliation{Institute of Modern Physics, Northwest University, Xian 710069, China}
\author{Wenli He}
%\email{hewenli@phy.nwu.edu.cn}
\affiliation{Institute of Modern Physics, Northwest University, Xian 710069, China}

%\date{\today}
\date{October 15,2001}


\begin{abstract}
A new approach for treating boundary Poisson structures based on causality and
locality analysis is proposed for a single scalar field with boundary
interaction. For the case of linear boundary condition, it is shown that the
usual canonical quantization can be applied systematically.
\end{abstract}

\maketitle

\vspace{0.5cm}

Quantization of classical field in the presence of various boundary conditions
is an old problem for which a systematical solution is still missing. This
problem is important because it is related to a vast range of physical
problems including, e.g. surface effect in condensed matter physics, cavity
QED, Casimir energy, two-dimensional integrability \cite{zf,dd,aa,bb},
mass generation \cite{9,10} and especially conformal field theory and
open string theory . Recently, there has been a renewed interests in this
problem among string theorists since the discovery of $D$-branes,
noncommutativity and other extended objects (like the so-called Horizontal
branes \cite{cc}) in string theory. In most recent papers on this subject, people are
tempted to use the Dirac procedure \cite{5} for constrained systems to treat the
inconsistency of Poisson structure with the boundary condition \cite{1,2,3,4}. However, for
at least several reasons we think that this treatment of boundary condition is
not quite appreciated. First, since the boundary condition is a constraint at
a specific spacial point (or a spacial hypersurface if there are more than one
spacial directions) which is of functional measure $0$ in the space of
field, the direct application of Dirac procedure necessarily fails because the
standard definition of Dirac Poisson brackets in this case would involve the
inverse of $\delta(0)$, which could hardly be given any practical sense. An
alternative way is to put the spacial direction with the boundary into a
lattice form and then implement the Dirac procedure. In this case the
construction of Dirac brackets seems to make perfect sense \cite{1,2,3,4}, but,
unfortunately, there seems to be no simple continuum limit for such lattice regularized
theories for which the Dirac brackets remain consistent. More over, the
appearance of an explicit length scale -- the lattice spacing -- is another
drawback of this formalism, which is especially unfavorable in scale invariant theories.

In this article, we shall propose a new method to treat the inconsistency
between the boundary condition and the canonical Poisson structure. Our
treatment is in fact a modified definition of the canonical structure
according to the analysis of the causality and locality of the theory in the
presence of boundary condition. For simplicity we shall consider only the
simplest case of a single scalar field in $(D+1)$-spacetime dimensions, where
the time direction and the first $D-1$ spacial directions are boundaryless,
and the only boundary condition appears in the direction of $x^{D}$, which
extends over $[0,+\infty)$.

\vspace{0.5cm}
%\special{bmp:d:/text/tex/fig1.bmp}
\begin{center}
\epsfbox{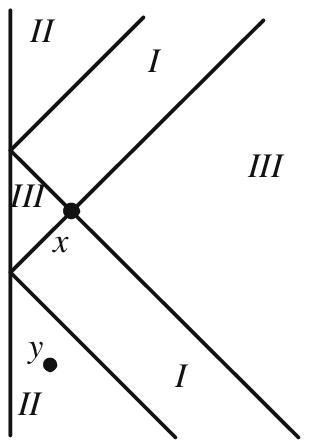}
{Figure 1: The lightcone}
\end{center}

Before going into the concrete action level analysis, let us first analyze the
causal structure of the scalar field theory in the presence of a boundary.
Figure 1 depicts the $(x^{0},x^{D})$ slice of the lightcone of the theory.
Noticing that the hypersurface $x^{D}=0$ is a reflecting barrier for light
signals, we have, in contrast to the case without boundaries, three different
zones which are denoted as zone $I$, $II$ and $III$ respectively. Zone $I$ and
$II$ are both lightlike, the difference is that zone $I$ is a reflectionless
zone which means that no signals of events happened in the causal past $y$ of
the observer at $x$ can be reflected from the boundary to the observer, while
events happened in zone $II$ can be reflected without breaking the causality.
Zone $III$ of the lightcone is the usual spacelike zone. Therefore, without
knowing any details of the action, we may conclude that the bare propagator of
the scalar field theory with such a lightcone must behave like%
\begin{align}
&  \Delta_{\mathcal{B}}(x,y)\nonumber\\
&  =\left\{
\begin{array}
[c]{ll}%
\Delta(x-y), & y\in\mathrm{zone}\quad I\\
\Delta(x-y)+\mathcal{B}\Delta(x-\sigma(y)), & y\in\mathrm{zone}\quad II\\
0, & y\in\mathrm{zone}\quad III
\end{array}
,\right.  \label{postulate}%
\end{align}
where $\Delta(x-y)$ is the standard propagator for the same theory in the
bulk, $\sigma(y)$ is the reflection image of $y$ with respect to the boundary,
i.e. if $y=(y^{0},y^{1},...,y^{D-1},y^{D})$, then $\sigma(y)=(y^{0}%
,y^{1},...,y^{D-1},-y^{D})$, and $\mathcal{B}$ is some operator which
represents the effectiveness of the boundary reflection. For ideal reflection,
we must have $||\mathcal{B}||=1$ (of cause the operator norm $||\quad||$ must
be assigned a proper sense -- we shall come back to this point later).

Now let us write down the action of a single scalar field $\varphi$ with a
bulk interaction $V(\varphi)$ and also a boundary interaction
$V_{B}(\varphi)$. It reads%
\begin{align*}
&  S[\varphi]\\
&  =\frac{1}{2}\int d^{D}x\int_{0}^{\infty}dx^{D}\left[  \partial_{M}%
\varphi\partial^{M}\varphi-m^{2}\varphi^{2}+2gV(\varphi)\right] \\
&  +\lambda_{B}\left.  \int d^{D}xV_{B}(\varphi)\right|  _{x^{D}=0},
\end{align*}
where throughout this article, $\int d^{D}x$ represents the integration over
all the transverse spacetime directions to the direction of $x^{D}$, the Roman
index $M$ runs from $0$ to $D$, whereas the
Greek index $\mu$ runs from $0$ to $D-1$. The constants $g,\lambda$
respectively represent the strength of the bulk and boundary couplings. As in
the case of ordinary field theories without boundaries, the canonical
conjugate momentum $\pi(x)$ is still defined via the bulk Lagrangian as%
\[
\pi(x)=\frac{\delta L}{\delta\partial_{0}\varphi(x)}=\partial_{0}\varphi(x).
\]

The variation of $S[\varphi]$ reads%
\begin{align*}
&  \delta S[\varphi]\\
&  =\int d^{D}x\int_{0}^{\infty}dx^{D}\\
&  \times\left[  \partial_{M}(\delta\varphi\partial^{M}\varphi)-\delta
\varphi\left(  \partial_{M}\partial^{M}\varphi+m^{2}\varphi-g\frac{\delta
V}{\delta\varphi}\right)  \right] \\
&  +\lambda\left.  \int d^{D}x\delta\varphi\frac{\delta V_{B}(\varphi)}%
{\delta\varphi}\right|  _{x^{D}=0}\\
&  =-\int d^{D}x\int_{0}^{\infty}dx^{D}\delta\varphi\left(  \partial
_{M}\partial^{M}\varphi+m^{2}\varphi-g\frac{\delta V}{\delta\varphi}\right) \\
&  +\left.  \int d^{D}x\delta\varphi\left[  \partial_{D}\varphi+\lambda
\frac{\delta V_{B}(\varphi)}{\delta\varphi}\right]  \right|  _{x^{D}=0}.
\end{align*}
Therefore, the condition $\delta S[\varphi]=0$ yields, for arbitrary
$\delta\varphi$, the following equation of motion and boundary condition,%
\[
\partial_{M}\partial^{M}\varphi+m^{2}\varphi-g\frac{\delta V}{\delta\varphi
}=0,
\]%
\begin{equation}
\left.  \partial_{D}\varphi+\lambda\frac{\delta V_{B}(\varphi)}{\delta\varphi
}=0\right|  _{x^{D}=0}. \label{BC}%
\end{equation}
The boundary condition (\ref{BC}) gives a constraint between $\partial
_{D}\varphi$ and $\frac{\delta V_{B}(\varphi)}{\delta\varphi}$ on the
spacetime hypersurface $x^{D}=0$, and thus the naive canonical Poisson
brackets%
\begin{align}
\{\varphi(x),\varphi(y)\}  &  =\{\pi(x),\pi(y)\}=0,\label{naivepb}\\
\{\varphi(x),\pi(y)\}  &  =\delta(x-y)\nonumber
\end{align}
(where $\delta(x-y)$ should be understood as $\delta^{(D)}(x-y)=\prod
_{i=1}^{D}\delta(x_{i}-y_{i})$) do not hold consistently.

As mentioned earlier, the usual Dirac procedure does not apply satisfactorily
to the case of boundary constraints without discretizing the spacial
coordinate with boundary condition. Fortunately, since now the inconsistency
only occur at the hypersurface $x^{D}=0$, we may expect, following the
principle of locality, that the modification to the naive Poisson brackets
should be nontrivial only on the same hypersurface. Moreover, since there is
no dependence on the canonical momentum in the boundary condition, only
$\{\varphi(x),\pi(y)\}$ needs to be modified. So, without loss of
generality, we assume that the correct $\{\varphi(x),\pi(y)\}$ take
the following form,%
\begin{equation}
\{\varphi(x),\pi(y)\}=\delta(x-y)+B(y)\delta(x-\sigma(y)), \label{Poisson}%
\end{equation}
where one should notice that $\delta(x-\sigma(y))$ is nonzero only if both
$x^{D}$ and $y^{D}$ are equal to $0$, and $B(y)$ is an operator acting on the
variable $y$ which represents the effect of boundary reflection. In this
article, we adopt a slightly modified definition for the $\delta$-function.
We assume that
\[
\int_{0}^{\infty}dx^{D}\delta(x^{D})=1,
\]
or, in terms of the standard definition for $\delta$-function, our
$\delta(x^{D})$ should be understood as $\lim_{\epsilon\rightarrow0^{+}}%
\delta(x^{D}+\epsilon)$.

Now let us check what form should the operator $B(y)$ take in order that
the new Poisson bracket (\ref{Poisson}) be consistent. For this purpose we
first write the boundary condition as a boundary constraint,%
\[
G=\int_{0}^{\infty}dx^{D}\delta(x^{D})\left[  \partial_{D}\varphi
+\lambda\frac{\delta V_{B}}{\delta\varphi}\right]  \simeq0.
\]
Examining the Poisson bracket $\{G,\pi(y)\}$ (we only need to do so because
$G$ naively Poisson commutes with $\varphi(y)$), one gets%
\begin{align*}
&  \{G,\pi(y)\}\\
&  =\int_{0}^{\infty}dx^{D}\delta(x^{D})\left(  \partial_{D}+\lambda
\frac{\delta^{2}V_{B}}{\delta\varphi^{2}}\right) \\
&  \times\left[  \delta(x-y)+B(y)\delta(x-\sigma(y))\right] \\
&  =\left[  B(y)\left(  \partial_{y^D}+\lambda\frac{\delta^{2}V_{B}}%
{\delta\varphi^{2}}\right)  -\left(  \partial_{y^D}-\lambda\frac{\delta^{2}%
V_{B}}{\delta\varphi^{2}}\right)  \right] \\
&  \times\delta^{(D-1)}(x-y)\delta(y^{D}).
\end{align*}
In the last equality, $\frac{\delta^{2}V_{B}}{\delta\varphi^{2}}$ is to
be regarded as a function of $y$. The consistence condition $\{G,\pi(y)\}=0$
yields%
\begin{equation}
B(y)=\frac{\partial_{y^D}-\lambda\frac{\delta^{2}V_{B}}{\delta\varphi^{2}}%
}{\partial_{y^D}+\lambda\frac{\delta^{2}V_{B}}{\delta\varphi^{2}}}. \label{BV}%
\end{equation}
The Poisson brackets (\ref{naivepb},\ref{Poisson}) with the operator $B(y)$
given as (\ref{BV}) then form a consistent set of Poisson structure for our
boundary scalar field theory. Noticing the fact that $B(y)$ always acts on
$\delta(x-\sigma(y))$, one may just denote $\partial_{y^D}$ as $\partial_D$.

The quantization of the above scalar field with boundary is still not an easy
task because one still needs to assign proper meaning to the operator ordering
and operator inverse appeared in the expression (\ref{BV}).

Fortunately, there is a simple illuminating case in which one does not need to
worry about the above problem, i.e. the case of linear boundary conditions. In
this special case, one simply take $V_{B}(\varphi)=-\frac{1}{2}\varphi^{2}$,
and thus the boundary interaction becomes just a boundary mass term. The
boundary condition (\ref{BC}) then becomes%
\begin{equation}
(\partial_{D}-\lambda)\varphi=0|_{x^{D}=0}, \label{BClinear}%
\end{equation}
and the operator $B(y)$ is now
\begin{equation}
B(y)=\frac{\partial_{D}+\lambda}{\partial_{D}-\lambda}. \label{BVlinear}%
\end{equation}
In the rest of this article, we shall be considering the scalar field theory
with this last boundary condition.

To quantize the theory with the boundary condition (\ref{BClinear}), one only
needs to quantize the corresponding free theory, i.e. the Klein-Gordon field
with the boundary (\ref{BClinear}) and then apply the standard perturbation
theory to introduce the effect of the quantized interaction term
$:V(\varphi):$ in the bulk.

Let us first write down the classical solution to the massive Klein-Gordon
equation obeying the boundary condition (\ref{BClinear}):%
\begin{align*}
\varphi(x)  &  =\int\frac{d^{D-1}k}{(2\pi)^{D-1}}\int_0^{+\infty}\frac{dk_D}{2\pi}\frac{1}{\sqrt
{\omega_{k}}}[a(k)f(k_D,x^{D})e^{-ik_{\mu}x^{\mu}}\\
&  +a^{\ast}(k)f^{\ast}(k_D,x^{D})e^{ik_{\mu}x^{\mu}}]|_{\omega_{k}=k_{0}},
\end{align*}
where $d^{D-1}k=dk_{1}...dk_{D-1}$. $k_D$ runs over $[0,+\infty)$
because the function
\[
f(k_D,x^{D})=e^{ik_{D}x^{D}}+B(-ik)e^{-ik_{D}x^{D}}
\]
with $B(-ik)$ being the Fourier image of our boundary operator $B(y)$,%
\[
B(-ik)=\frac{ik_{D}-\lambda}{ik_{D}+\lambda},
\]
is complete over the half $k_D$ line,
\begin{align}
&  \int_0^{+\infty}\frac{dk^{D}}{2\pi}f(k_D,x^{D})f^{\ast}(k_D,y^{D})\nonumber\\
&  =\delta(x^{D}-y^{D})+B(y)\delta(x^{D}+y^{D}). \label{forth}%
\end{align}
Notice that $f(k_D,x^{D})$ solves the equation $(\partial_{D}-\lambda)f(k_D,x^{D})=0|_{x^D=0}$.
The $(D+1)$-momentum $k$ naturally satisfies the standard mass shell condition%
\[
k^{2}-m^{2}=0.
\]

The canonical quantization for the Klein-Gordon field is now accomplished by
replacing the Poisson bracket $\{\quad,\quad\}$ by the equal-time commutator
$-i[\quad,\quad]$,%
\begin{align}
\lbrack\varphi(x),\varphi(y)]  &  =[\pi(x),\pi(y)]=0,\nonumber\\
\lbrack\varphi(x),\pi(y)]  &  =i\left[  \delta(x-y)+B(y)\delta(x-\sigma
(y))\right]  . \label{commu}%
\end{align}
Let us remind that two special well-known cases are already contained in this
simple illustration, namely, the Neumann boundary condition (which corresponds
to $\lambda=0$ or $B=1$) and Dirichlet boundary condition (for which
$\lambda=\infty$ or $B=-1$). Our result (\ref{commu}) agrees with the known
result \cite{1} on these two special cases. For generic value of $\lambda$,
the relation $B(-ik)B^{\dagger}(-ik)=1$ gives a simple
explanation to the condition $||\mathcal{B}||=1$ mentioned earlier.

Now, going to the momentum space representation, one simply replaces the
momentum space coefficients $a(k),a^{\ast}(k)$ by their corresponding
operators $\hat{a}(k)$ and $\hat{a}^{\dagger}(k)$,%
\begin{align*}
\varphi(x)  &  =\int\frac{d^{D-1}k}{(2\pi)^{D-1}}\int_0^{+\infty}\frac{dk_D}{2\pi}\frac{1}{\sqrt
{\omega_{k}}}[\hat{a}(k)f(k_D,x^{D})e^{-ik_{\mu}x^{\mu}}\\
&  +\hat{a}^{\dagger}(k)f^{\ast}(k_D,x^{D})e^{ik_{\mu}x^{\mu}}]|_{\omega
_{k}=k_{0}}.
\end{align*}
Using this last expression and $\pi(x)=\partial_{0}\varphi(x)$, one can get
the commutation relation for the momentum space operators
$\hat{a}(k),\hat{a}^{\dagger}(k)$,%
\begin{align*}
\lbrack\hat{a}(k),\hat{a}(k^{\prime})]  &  =[\hat{a}^{\dagger}%
(k),\hat{a}^{\dagger}(k^{\prime})]=0,\\
\lbrack\hat{a}(k),\hat{a}^{\dagger}(k^{\prime})]  &  =(2\pi)^{D}%
\delta(k-k^{\prime}),
\end{align*}
with $k_D, k'_D \geq 0$. This is the Heisenberg type algebra arisen in usual field
theory restricted to the half momentum space $k_D\geq 0$. Therefore, we can
use the Fock space for ordinary Klein-Gordon field with the same restriction
as the space of states for our theory. In
particular, we have the vacuum state $|0\rangle$ which is annihilated by
$\hat{a}(k)$ and we choose the normalization for single particle states to be%
\[
\langle0|\hat{a}(k)\hat{a}^{\dagger}(k^{\prime})|0\rangle=(2\pi)^{D}%
\delta(k-k^{\prime}).
\]

Following the standard argument in quantum field theory (see, e.g. \cite{11},
we now can evaluate the propagator
\begin{align*}
D^{(B)}(x,y)  &  \equiv\langle0|T[\varphi(x)\varphi(y)]|0\rangle\\
&  =\theta(x^{0}-y^{0})\langle0|\varphi(x)\varphi(y)|0\rangle\\
&  +\theta(y^{0}-x^{0})\langle0|\varphi(y)\varphi(x)|0\rangle,
\end{align*}
which turns out to be%
\begin{align*}
D^{(B)}(x,y)  &  =\int\frac{d^{D}k}{(2\pi)^{D}}\int_0^{+\infty}\frac{dk_D}{2\pi}\\
&  \times\frac{i}{k^{2}-m^{2}+i\epsilon}f(k_D,x^{D})f^{\ast}(k_D,y^{D}%
)e^{-ik_{\mu}(x-y)^{\mu}},
\end{align*}
where the integration with respect to $k_{0}$ is a contour integration over
the complex $k_{0}$-plane which runs from $-\infty$ below the real axis to
$k_{0}=0$, crossing the real axis and goes to $+\infty$ above the real axis.
Substituting the definitions of $f(k,x^{D}),f^{\ast}(k,y^{D})$ into the last
equation and taking into account the relation (\ref{forth}), we get%
\begin{equation}
D^{(B)}(x,y)=D_{F}(x-y)+B(y)D_{F}(x-\sigma(y)), \label{propagator}%
\end{equation}
where
\[
D_{F}(x-y)=\int\frac{d^{D+1}k}{(2\pi)^{D+1}}\frac{i}{k^{2}-m^{2}+i\epsilon
}e^{-ik_{M}(x-y)^{M}}%
\]
is the standard Feynman propagator in $D+1$ dimensions.

Notice that, following the standard discussion on the causality properties of
the Feynman propagator \cite{11}, we would find that the result (\ref{propagator})
agrees perfectly with our early postulation (\ref{postulate}). The propagator
(\ref{propagator}) is then the very basic object in the Feynman rules of
perturbation theory for the interacting boundary scalar field theory with
generic interacting potential $V(\varphi)$ in the bulk.

We may also evaluate the retarded Green's function for the boundary
Klein-Gordon field as well. After some simple calculations, we have%
\begin{align*}
D_{R}^{(B)}(x,y)  &  \equiv\theta(x^{0}-y^{0})\langle0|[\varphi(x),\varphi
(y)]|0\rangle\\
&  =D_{R}(x-y)+B(y)D_{R}(x-\sigma(y)),
\end{align*}%
\[
D_{R}(x-y)=\int\frac{d^{D+1}k}{(2\pi)^{D+1}}\frac{i}{k^{2}-m^{2}}%
e^{-ik_{M}(x-y)^{M}},
\]
where the integration contour for $k_{0}$ is taken to be running from
$-\infty$ to $+\infty$ above the real axis. The fact that $D_{R}^{(B)}(x,y)$
is a green's function is now assured by the following equation,%
\[
(\partial_{M}\partial^{M}+m^{2})D_{R}^{(B)}(x,y)=\delta(x-y)+B(y)\delta
(x-\sigma(y)).
\]

Before finishing this article, let us make some more comments on the commutation
relation (\ref{commu}). Though by definition we know that the quantities
$\varphi(x), \pi(x)$ live only on the half space $x^D \geq 0$, we may, however,
imagine to analytic continue them to the whole spacetime. In that case, writing
\[
\varphi(x)=\int \frac{d^D \mathbf{k}}{(2\pi)^D} A(\mathbf{k}) e^{ikx},
\hspace{0.2cm}
\pi(x)=\int \frac{d^D \mathbf{k}}{(2\pi)^D} A^\dagger(\mathbf{k}) e^{-ikx},
\]
where $d^D\mathbf{k}=dk_1...dk_D$ with $k_D$ running from $-\infty$ to $+\infty$,
we have, for the momentum space operators $A(\mathbf{k})$ and $A^\dagger(\mathbf{k})$, the following
commutation relations,
\[
\lbrack A(\mathbf{k}), A(\mathbf{k}')\rbrack =
\lbrack A^\dagger(\mathbf{k}), A^\dagger(\mathbf{k}')\rbrack = 0,
\]
\[
\lbrack A(\mathbf{k}), A^\dagger(\mathbf{k}')\rbrack = \delta(\mathbf{k}-\mathbf{k}') +
\frac{ik'_D-\lambda}{ik'_D+\lambda} \delta(\mathbf{k}-\sigma(\mathbf{k}')).
\]
These relations are the simplest case of the boundary exchange algebra introduced
in \cite{6} and further explored in \cite{7}.

So far, in this article, we considered the problem of boundary Poisson
structure for the case of a single scalar field theory with boundary
interaction $V_{B}(\varphi)$. The fact that the consistent Poisson bracket
between $\varphi(x)$ and $\pi(y)$ depends on the second variation of $V_{B}$
with respect to $\varphi$ makes the problem of canonical quantization in the
presence of boundary a difficult task for generic $V_{B}$. For linear boundary
conditions, i.e. for $V_{B}$ at most quadratic in $\varphi$, the canonical
quantization can be pursued without difficulties. Though we have obtained the
consistent Poisson structure for generic boundary interaction $V_{B}$
depending only on the field $\varphi$, we did not include more generalities,
i.e. the cases in which $V_{B}$ depends also on spacetime derivatives of the
field $\varphi$, and the cases in which there are more than one scalar fields
and $V_{B}$ couples different components of the fields etc.
It is also tempting to study the case when the fundamental field is is group-valued,
like the principal chiral model studied in \cite{dd}. The direct quantization for
generic $V_B$ is also a fascinating subject (some progress for the case of sine-Gordon
field with integrable boundary condition has already been made in \cite{aa,bb}).
%We wish to publish
%the detailed study of these more complicated cases elsewhere.

We thank M. Mintchev for useful comment on the manuscript.
This work is supported by the National Natural Science Foundation of china.

\vspace{0.3cm}

%\bibliography{}

\end{document}